\documentclass[sigconf]{acmart} % arXiv
\AtBeginDocument{%
  \providecommand\BibTeX{{%
    \normalfont B\kern-0.5em{\scshape i\kern-0.25em b}\kern-0.8em\TeX}}}

\setcopyright{acmlicensed}
\copyrightyear{2024}
\acmYear{2024}
\acmDOI{10.1145/3589335.3651911}

\acmConference[WWW '24 Companion]{Companion Proceedings of the ACM Web Conference 2024}{May 13--17, 2024}{Singapore, Singapore}
\acmBooktitle{Companion Proceedings of the ACM Web Conference 2024 (WWW '24 Companion), May 13--17, 2024, Singapore, Singapore}
\acmISBN{979-8-4007-0172-6/24/05}
\begin{document}

\title[A Bayesian Framework for Measuring Association]{A Bayesian Framework for Measuring Association and Its Application to Emotional Dynamics in Web Discourse}
\author{Henrique S. Xavier}
\email{hxavier@nic.br}
\orcid{0000-0002-9601-601X}
\affiliation{%
  \institution{NIC.br}
  \department{Ceweb.br}
  \streetaddress{Av. das Nações Unidas, 11541, 7º andar}
  \city{São Paulo}
  \state{SP}
  \country{Brazil}
  \postcode{04578-000}
}

\author{Diogo Cortiz}
\email{diogo@nic.br}
\orcid{0000-0002-5875-8602}
\affiliation{%
  \institution{NIC.br}
  \department{Ceweb.br}
  \streetaddress{Av. das Nações Unidas, 11541, 7º andar}
  \city{São Paulo}
  \state{SP}
  \country{Brazil}
  \postcode{04578-000}
}

\author{Mateus Silvestrin}
\orcid{0000-0002-3482-3676}
\affiliation{%
  \institution{Mackenzie Presbyterian University}
  \streetaddress{Rua da Consolação, 930}
  \city{São Paulo}
  \state{SP}
  \country{Brazil}
  \postcode{01302-907}
}

\author{Ana Luísa Freitas}
\orcid{0000-0002-9383-2679}
\affiliation{%
  \institution{Mackenzie Presbyterian University}
  \streetaddress{Rua da Consolação, 930}
  \city{São Paulo}
  \state{SP}
  \country{Brazil}
  \postcode{01302-907}
}

\author{Letícia Yumi Nakao Morello}
\orcid{0000-0002-3053-9299}
\affiliation{%
  \institution{Mackenzie Presbyterian University}
  \streetaddress{Rua da Consolação, 930}
  \city{São Paulo}
  \state{SP}
  \country{Brazil}
  \postcode{01302-907}
}

\author{Fernanda Naomi Pantaleão}
\orcid{0000-0003-1038-4370}
\affiliation{%
  \institution{Mackenzie Presbyterian University}
  \streetaddress{Rua da Consolação, 930}
  \city{São Paulo}
  \state{SP}
  \country{Brazil}
  \postcode{01302-907}
}

\author{Gabriel Gaudencio do Rêgo}
\orcid{0000-0003-3304-4723}
\affiliation{%
  \institution{Mackenzie Presbyterian University}
  \streetaddress{Rua da Consolação, 930}
  \city{São Paulo}
  \state{SP}
  \country{Brazil}
  \postcode{01302-907}
}

\renewcommand{\shortauthors}{Xavier et al.}

\begin{abstract}
This paper introduces a Bayesian framework designed to measure the degree of association between categorical random variables. The method is grounded in the formal definition of variable independence and is implemented using Markov Chain Monte Carlo (MCMC) techniques. Unlike commonly employed techniques in Association Rule Learning, this approach enables a clear and precise estimation of confidence intervals and the statistical significance of the measured degree of association. We applied the method to non-exclusive emotions identified by annotators in 4,613 tweets written in Portuguese. This analysis revealed pairs of emotions that exhibit associations and mutually opposed pairs. Moreover, the method identifies hierarchical relations between categories, a feature observed in our data, and is utilized to cluster emotions into basic-level groups.
\end{abstract}




%  \includegraphics[width=\textwidth]{images/dalle_smiley_network_teaser}
%  \caption{A painting expressing the associations between sentiments.}
%  \Description{A painting of smiley faces connected by lines as nodes in a network.}

\received{5 February 2024}
%\received[revised]{12 March 2009} % arXiv
%\received[accepted]{5 June 2009}  % arXiv

\maketitle

\section{Introduction}
\label{sec:intro}

A considerable amount of content is generated on the Web every day. Individuals and organizations upload videos and images on various platforms, publish articles on news websites and blogs, create posts on social media, edit content on wikis, and leave comments on many of these sites. This data is often analyzed for commercial and scientific purposes, and in this process, it is common to associate categorical features with it. For example, images and videos may undergo detection for objects, copyright violations, deepfakes and offensive content. At the same time, textual data can be classified into various sentiments, moral categories, topics and languages, and can be tested for toxicity, plagiarism, factual accuracy and vandalism, among other things. All these processes result in categorical features that are ascribed to the content.

Understanding whether features are associated, independent, or opposed may be relevant in these contexts. Associated features tend to co-occur due to underlying relations between them (e.g., the detection of a motorbike in a picture and the detection of a helmet), while observed co-occurrences for independent features are incidental (e.g., the numbers facing up on a two-dice roll). Lastly, two features are opposed if the occurrence of one reduces the chance of the other occurring. The analysis of these relations is commonly known as Association Rule Learning \cite{Agrawal1993}. Its application helps in comprehending whether a given topic (e.g., politics, sports, environment, science, humor), factuality level, or political stance is associated with a specific set of emotions, whether certain descriptors accompany an entity recognized in a text, or whether two types of objects can be considered opposing classes or if they can both appear in an image.

Traditional Association Rule Learning methods rely on point estimations of some measure of ``interestingness'' \cite{Agrawal1993, Brin1997, Omiecinski2003, Tan2004}. This characteristic is likely connected to the fact that most methods aim to test associations among a vast number of variables (e.g., thousands of products available in a supermarket), and point estimations are computationally economical. Speed becomes a constraint as the number of potential pairwise associations grows quadratically with the number of variables, and the number of associations between larger combinations grows even more rapidly. Moreover, these methods are tailored for situations where the available data contains numerous instances (e.g., many thousands of purchases made by customers); in large samples, point estimations are reasonably precise.

However, point estimates on their own lack information about the statistical significance of the association and are unaccompanied by confidence intervals. When such assessments are of interest, resorting to separate statistical tests becomes necessary, and the most well-known tests might not align with the experimental setup typically associated with data collected on the Web. In contrast, if one can compute the full posterior probability distribution for some measure of independence between two variables, then all the information about their association available in the data can be extracted. This additional information can be beneficial when dealing with a few instances. 

Focusing on extracting all available information from the data, we present a Bayesian framework for assessing the degree of association (or opposition) between categorical random variables. In our framework, detailed in Section \ref{sec:method}, the measure of ``interestingness'' is defined by the deviation from the definition of variable independence, known in the literature as \emph{added value} \cite{Tan2004}. The method yields an estimation of the posterior probability distribution for the measure, enabling the computation of point estimates, confidence intervals, and the assessment of statistical significance within a unified and theoretically grounded framework. Due to decreased speed compared to traditional methods, the framework is tailored for datasets with fewer features. To enhance its usability, we have released BRASS, an open-source Python implementation of the method\footnote{\url{https://github.com/cewebbr/brass}}.

After confirming the functionality of our method and its implementation on synthetic data (refer to Section \ref{sec:simulations}), we showcase the results derived from this framework by applying it to 30 emotional categories annotated on 4,613 posts on the X platform (known as Twitter at the time of data collection) in Section \ref{sec:application}. In this application, we gauged the association between emotions present in the same post and compared our findings to expectations derived from prominent models in emotional research. The evaluation of $30 \cdot 29 / 2 = 435$ pairwise associations can be accomplished within a few hours on a modern laptop.

\section{Previous and related work}
\label{sec:related-work}

Previous efforts in identifying associations between categorical random variables, as detailed in Section \ref{sec:apriori}, typically do not assess the statistical significance of their findings. Therefore, we summarize standard methods for testing statistical significance in Section \ref{sec:significance}.

\subsection{Identifying interesting associations}
\label{sec:apriori}

Arguably the most renowned paper on Association Rule Learning is by Agrawal, Imielinski, and Swami \cite{Agrawal1993}. The paper revolves around discovering interesting associations among many variables, such as item sets purchased in a retail store. The objective is to determine if a set of items, termed \emph{antecedent}, is frequently accompanied by another set of items, known as \emph{consequent}. In this context, the paper focused on diminishing the computational cost of testing all possible associations between item sets.

One approach to achieve this is by disregarding candidate associations whose \emph{support} -- i.e., the fraction of purchases containing both the antecedent and consequent item sets -- falls below a specified threshold. The rationale is that combinations of item sets with higher support co-occur more frequently, thus indicating a more substantial chance of association. An advantage of using support as a preliminary filter lies in its anti-monotonicity, also known as the downward-closure property: subsets created with items from a frequent set are also frequent. This property reduces the number of required tests and expedites the analysis.

Following this initial selection, a second measure of interestingness is computed, as support alone is not sufficient to identify associated item sets. In \cite{Agrawal1993}, a measure called \emph{confidence} is employed, though various other measures have been proposed and compared \cite{Brin1997, Sahar1999, Tan2004}. Candidates surpassing a specified threshold on this measure are ultimately considered interesting.

One drawback of approaches relying on minimum support is their inherent incapacity to identify opposing variables. Since opposing variables appear together less frequently than one would expect by chance, they exhibit low support. However, beyond that, the methods mentioned above typically treat observed frequencies as equivalents to probabilities. As explained by Sahar and Mansour, ``Note that we use the term probability here rather loosely. $Pr[C]$ denotes the fraction of entries in the training set for which $C$ holds. We assume that all our samples are large enough to ensure that the frequency counts are accurately indicative of the actual probabilities'' \cite{Sahar1999}. This assumption also holds for papers employing Bayesian Networks for Association Rule Learning \cite{Tian2013,Kharya2019}, despite the term ``Bayesian.''

In reality, observed frequencies serve as point estimates for the probability of the underlying stochastic process generating the data. If the experiment were replicated under identical conditions, it would be improbable to observe precisely the same frequency. Generally, the research interest lies not in patterns from a single experiment but in patterns of the underlying process. Treating observed frequencies as probabilities overlooks statistical fluctuations and uncertainties in any derived measure of interestingness. Consequently, previous methods are primarily suitable for categories (or items) that manifest frequently in the data, as articulated by Sahar and Mansour.

An exception to this is the work of DuMouchel \cite{DuMouchel1999}, which clearly distinguishes the properties of the data and those of its generating process. The paper assumes that the number $N_{ij}$ of co-occurrences of two events $i$ and $j$ is drawn from a Poisson distribution with an unknown mean rate $\mu_{ij}$. It then employs Bayes' Theorem to compute the posterior distribution for $\lambda_{ij} \equiv \mu_{ij} / E_{ij}$, where $E_{ij}$ is the expected rate of co-occurrences under the hypothesis of statistical independence between $i$ and $j$.

Modeling co-occurrences with a Poisson distribution is appropriate when the number of trials is unknown. If we choose a dataset with a specified number of instances and aim to determine the probability at which certain features appear together, the binomial or multinomial distributions are more suitable models. This is often the scenario in analyses conducted over Web data. Nonetheless, applications of Association Rule Learning to Web data, such as constructing recommendations from clickstream data and investigating follow patterns on social media, still rely on point estimates of probabilities \cite{Mobasher2001,Kruse2022}.

\subsection{Tests of statistical significance}
\label{sec:significance}

In the literature, statistical significance tests for associations between categorical random variables typically emerge in settings where both the sample sizes and the number of candidate associations are small. These tests are usually based on contingency tables that organizes the results from a collection of independent measurements of categorical variables, such as sentiment annotations on comments made on a web platform. These tables provides a concise overview of the number of instances within the sample that exhibit each possible combination of categories (refer to Table \ref{tab:contingency-example} for an example). This table's final row and column display the overall totals for each category, often referred to as margins.

\begin{table}[ht]
  \caption{An example of a contingency table illustrating the detection of two sentiments (joy and admiration) in our corpus.}
  \label{tab:contingency-example}
  \begin{tabular}{l|cc|r}
                  & No Joy & Joy & Total\\
    \midrule
    No Admiration & 4274 & 112 & 4386 \\
    Admiration    & 205 & 22 & 227 \\
    \midrule
    Total         & 4479 & 134 & 4613 \\
  \end{tabular}
\end{table}

Several statistical tests are designed to identify associations between categorical variables. One such test is Fisher's exact test, which assumes the null hypothesis of variable independence and assesses it using $p$-values. This test considers that both margins of the contingency table (comprising the total number of occurrences for every category) are known \cite{Fisher1922}. Consequently, the data is modeled by a hypergeometric distribution.

On the other hand, Barnard's and Boschloo's tests are similar to Fisher's, with the key distinction being that only one margin is known. This margin corresponds to the total number of occurrences for each category within one of the variables \cite{Barnard1945, Boschloo1970}. In such cases, the underlying distribution is the product of two binomial distributions.

Unlike the tests above, our proposed test does not require prior knowledge of the margins of contingency tables apart from the total number of instances (the number 4,613 in the bottom-right corner of Table \ref{tab:contingency-example}). Consequently, we adopt a multinomial distribution to model the numbers within the table. This approach is particularly suited for analysis of samples taken at random from the Web, such as sentiment detection in posts from social media. Additionally, our method does not rely on assuming a null hypothesis (i.e., the supposition that there are no relations between the variables). Instead, we employ Bayes' theorem to estimate the underlying posterior for the parameters of the multinomial distribution, specifically the success probabilities for each possible outcome.

While tests like Fisher's exact test can identify dependencies among variables, they do not inherently quantify the strength of the detected dependence, i.e., the extent of association. Measures of this association strength are typically provided by statistics such as the Odds ratio, Tetrachoric correlation, Goodman and Kruskal's lambda, or the Uncertainty coefficient \cite{Szumilas2010, Olsson1979, Goodman1954, Press2007}. In contrast, by computing a posterior distribution for the degree of association, our method integrates the detection of dependence and the measurement of the degree of association into a single tool. An additional benefit of our approach is that it allows for estimating the degree's uncertainty.  

\section{The method}
\label{sec:method}

In a nutshell, our method consists of: modeling the observed data as a draw from a multinomial distribution; calculating the likelihood of the data given this distribution; employing Bayes' Theorem and an appropriate prior to derive the posterior distribution for the parameters of the multinomial distribution; and deriving a distribution for a measure of variable independence from these parameters. 

\subsection{Setting the stage}

Categorical variables with more than two allowed values can be transformed into binary dummy variables through one-hot encoding. This transformation is illustrated by the variable $V_1$ in Table \ref{tab:dataset-example} and the columns X, Y, and Z. Moreover, binary variables can also represent item sets with more than one item, as shown in the last column of Table \ref{tab:dataset-example} for "items" X and $V_2$. Therefore, we will focus the derivation of our method on the relationship between two binary variables, whose values can be 0 or 1. In this scenario, for each instance of a pair of dummy variables $A$ and $B$, there are four possible outcomes: $(A, B)$ can be $(0, 0)$, $(0, 1)$, $(1, 0)$, or $(1, 1)$.

\begin{table}[ht]
  \caption{A toy example of a dataset with 6 instances (identified by the ``ID'' column) and 2 categorical variables: $V_1$ (with 3 possible values) and $V_2$ (with two possible values). Columns 
  X, Y and Z form a one-hot encoding for $V_1$.}
  \label{tab:dataset-example}
  \begin{tabular}{ccccccc}
    ID & $V_1$ & X & Y & Z & $V_2$ & $(X \wedge V_2)$ \\
    \midrule
     1 & Y & 0 & 1 & 0 & 1 & 0 \\
     2 & Z & 0 & 0 & 1 & 0 & 0 \\
     3 & X & 1 & 0 & 0 & 1 & 1 \\
     4 & X & 1 & 0 & 0 & 0 & 0 \\
     5 & Y & 0 & 1 & 0 & 0 & 0 \\
     6 & X & 1 & 0 & 0 & 1 & 1 \\
  \end{tabular}
\end{table}

\subsection{Derivation}
\label{sec:derivation}

In a sample of size $N$, we can count the times each possible outcome occurred. We refer to these counts as $N_{00}$, $N_{01}$, $N_{10}$ and $N_{11}$, respectively, and we have the constraint $N=\sum_{ij}N_{ij}$ for $i=0,1$ and $j=0,1$. If the instances in the sample were randomly generated through a stationary process and their outcomes are independent, then the counts follow a multinomial distribution:

\begin{equation}
 P (\boldsymbol{N}|N,\boldsymbol{p}) = 
 \frac{N!}{N_{00}!N_{01}!N_{10}!N_{11}!} 
 p_{00}^{N_{00}}p_{01}^{N_{01}}p_{10}^{N_{10}}p_{11}^{N_{11}},
 \label{eq:likelihood}
\end{equation}
where $P(\ldots)$ above is the probability mass function of the multinomial distribution, $\boldsymbol{N} \equiv (N_{00},\, N_{01},\, N_{10},\, N_{11})$, $p_{ij}$ represents the probability of the configuration $(i, j)$ to occur for one instance, and $\boldsymbol{p} \equiv (p_{00},\, p_{01},\, p_{10},\, p_{11})$.

We can use Bayes' theorem to compute the posterior probability density function for $\boldsymbol{p}$:

\begin{equation}
\begin{split}
 P'(\boldsymbol{p}|N,\boldsymbol{N})\mathrm{d}^4p = 
 \frac{P'(\boldsymbol{p}|N)\mathrm{d}^4p}{P(\boldsymbol{N}|N)}
 P(\boldsymbol{N}|N,\boldsymbol{p}).
\end{split}
\label{eq:posterior}
\end{equation}
In the equation above, $\mathrm{d}^4p\equiv \prod_{ij}\mathrm{d}p_{ij}$ is the differential volume in the $\boldsymbol{p}$ parameter space, while $P'(\boldsymbol{p}|N)$ represents the prior probability density distribution for $\boldsymbol{p}$.
Assuming that the prior is independent of $N$, we can remove the latter from the conditional, yielding $P'(\boldsymbol{p}|N) = P'(\boldsymbol{p})$.

We have chosen to use a uniform prior while imposing the constraint $\sum_{ij}p_{ij} = 1$. This results in a Dirichlet distribution with $\alpha_{ij}=1$, often called the flat Dirichlet distribution. For arbitrary positive integers $\boldsymbol{\alpha} = (\alpha_{00},\, \alpha_{01},\, \alpha_{10},\, \alpha_{11})$, the Dirichlet distribution is given by: 

\begin{equation}
    D(\boldsymbol{p}; \boldsymbol{\alpha}) \equiv 
    \frac{(\sum_{ij} \alpha_{ij} - 1)!}{\prod_{ij}(\alpha_{ij} - 1)!}
    \prod_{ij} p_{ij}^{\alpha_{ij} - 1}.
\end{equation}

It is helpful to note that the Dirichlet distribution is the conjugate prior of the multinomial distribution, meaning that the posterior distribution resulting from a multinomial likelihood and a Dirichlet prior is also a Dirichlet distribution:

\begin{equation}
    P'(\boldsymbol{p}|N,\boldsymbol{N}) = 
    D(\boldsymbol{p}; \boldsymbol{N} +\boldsymbol{\alpha}).
\end{equation}
Given that we have chosen $\alpha_{ij}=1$, the equation above leads to:

\begin{equation}
    P'(\boldsymbol{p}|N,\boldsymbol{N}) = 
    \frac{(N+ 3)!}{N_{00}!N_{01}!N_{10}!N_{11}!} 
 p_{00}^{N_{00}}p_{01}^{N_{01}}p_{10}^{N_{10}}p_{11}^{N_{11}},
 \label{eq:posterior-final}
\end{equation}
under the constraint $\sum_{ij}p_{ij}=1$. Aside from a normalization factor, this is equivalent to the right-hand side of Eq. \ref{eq:likelihood}. Such similarity is expected, as the two expressions differ only by the normalization factor $P(\boldsymbol{N}|N)$ and a uniform prior.

Eq. \ref{eq:posterior-final} represents our posterior probability density distribution for $\boldsymbol{p}$ -- our best understanding, given the data $\boldsymbol{N}$, of the probabilities of the underlying process producing $(i, j)$ for the variable pair $(A, B)$. However, our focus is not on the individual $p_{ij}$ values but on a measure of dependence between $A$ and $B$ derivable from $p_{ij}$. 

The definition of independence for two random variables $A$ and $B$ is typically expressed as $P(A=a|B=b)=P(A=a)$ for all values of $a$ and $b$ \cite{Pearl2016}. In the context of binary variables, it is sufficient to have $P(A=1|B=1)=P(A=1)$. Hence, our focus is on the quantity:

\begin{equation}
\Delta P(A,B) \equiv P(A=1|B=1) - P(A=1), 
\label{eq:deltaP}
\end{equation}
which is called \emph{added value} \cite{Tan2004} and serves as our measure of association. It quantifies how much the probability of $A$ occurring increases when $B$ is identified. 

From the posterior distribution of $\Delta P(A, B)$ (or $\Delta P$ for brevity), we can:

\begin{itemize}
\item determine if the variables are associated ($\Delta P > 0$) or opposed ($\Delta P < 0$);
\item calculate the expected value $\langle \Delta P \rangle$, quantifying the strength of dependence;
\item establish a confidence interval for $\Delta P$, including its standard deviation; and
\item test the hypothesis of independence between $A$ and $B$ by computing the $p$-value for $\Delta P=0$. For independent variables, this posterior should exhibit a significant probability past 0, while for dependent variables, we should be confident that they are either associated or opposed.
\end{itemize}
To compute $\Delta P(A,B)$ from $p_{ij}$ values, it is important to note that:
\begin{equation}
    P(A=1) = p_{10} + p_{11}
\end{equation}
and
\begin{equation}
    P(A=1|B=1) = \frac{p_{11}}{p_{01} + p_{11}}.
\end{equation}
Therefore, we have that:
\begin{equation}
\Delta P(A,B) = \frac{p_{11}}{p_{01} + p_{11}} - p_{10} - p_{11}. 
\label{eq:deltaP-final}
\end{equation}

The posterior for $\Delta P$ can be computed from Eq. \ref{eq:posterior-final} by substituting the variables $p_{ij}$ with a set that includes $\Delta P$ and then marginalizing over the remaining variables while ensuring they satisfy the constraint $\sum_{ij}p_{ij} = 1$. Unfortunately, it is not clear that the associated integrals can be performed analytically, so we resorted to numerical methods, described in Section \ref{sec:implementation}, as part of our implementation.

Equation \ref{eq:deltaP} shows that $\Delta P(A, B)$ is asymmetric under the permutation of $A$ and $B$, meaning that the added values (or probability boosts) $\Delta P(A, B)$ and $\Delta P(B, A)$ are generally distinct. As we will observe in Section \ref{sec:application}, this asymmetry can reveal hierarchical relationships between the two variables.

Despite this asymmetry, applying Bayes' theorem to Eq. \ref{eq:deltaP} leads to: 

\begin{equation}
\Delta P(B,A) = \frac{P(B=1)}{P(A=1)} \Delta P(A,B). 
\label{eq:permutation}
\end{equation}
Given that probabilities are always positive, this equation implies that $\Delta P(A, B)$ and $\Delta P(B, A)$ always share the same sign. Consequently, we can test for independence between $A$ and $B$ using either $\Delta P(A, B)$ or $\Delta P(B, A)$, as the posterior probability mass crossing zero is identical. This exchangeability is a good sign, as having $A$ associated with $B$ while $B$ is not associated with $A$ would be inconsistent.

\subsection{Implementation}
\label{sec:implementation}

We developed an open-source Python module called \texttt{brass} that estimates the posterior probability density distribution for $\Delta P$, along with ancillary quantities such as $p_{ij}$, $P(A=1)$, and $P(A=1|B=1)$, given the input data and the prior parameter $\boldsymbol{\alpha}$. The input data, representing observations of two binary random variables, can be provided as a table of instances or as elements of a contingency table. The main code follows these steps:

\begin{enumerate}
    \item Transform the input data into a contingency table if it was provided as a table of instances (rows) and variables (columns).
    \item Set up an internal representation of the posterior $D(\boldsymbol{p}; \boldsymbol{N} + \boldsymbol{\alpha})$; when $\alpha_{ij}=1$, this corresponds to Eq. \ref{eq:posterior-final}.
    \item Sample the posterior using the Markov Chain Monte Carlo method (MCMC) \cite{Press2007}, implemented in the PyMC package\footnote{\url{https://www.pymc.io}} \cite{Salvatier2016}.
    \item Compute $\Delta P$ from the sampled $\boldsymbol{p}$ using Eq. \ref{eq:deltaP-final}, along with other quantities of interest, and output the results in a Pandas\footnote{\url{https://pandas.pydata.org}} DataFrame.
\end{enumerate}
The \texttt{brass} module includes functions for computing the expected value, standard deviation, and $p$-values for $\Delta P$, along with capabilities for plotting marginal distributions. Additionally, it offers methods to generate synthetic pairs of binary variables based on specified values for $P(A)$, $P(B)$, and $\Delta P(A, B)$.

We chose to employ MCMC for sampling the posterior as it simplifies the enforcement of the constraint $\sum_{ij}p_{ij}=1$ and provides marginal posterior distributions for any quantity derived from $p_{ij}$, including $\Delta P$. In essence, MCMC randomly selects points in the $\boldsymbol{p}$ space in a manner that their frequency is proportional to the value of the function it samples; in our case, $D(\boldsymbol{p}; \boldsymbol{N} + \boldsymbol{\alpha})$. In other words, MCMC generates a sample of the population described by this distribution. All calculations that can be performed with other samples can be executed similarly.

The implementation of our method and all the analyses presented in this article are available at: 
\url{https://github.com/cewebbr/brass}.
%\url{https://example.org/anonymized}.

\subsection{Demonstration on simulations}
\label{sec:simulations}

To test our method and implementation, we first generated eight distinct synthetic datasets of pairs of binary variables, each with a different $\Delta P(A, B)$, $P(A=1)$, $P(B=1)$, and sample size $N$. These synthetic pairs were created by initially sampling $B$ according to $P(B=1)$ and then using conditional probabilities for $A$, derived from $\Delta P(A, B)$, to sample $A$ based on the observed values of $B$ for each instance. 

We then applied our method to the synthetic data. When sampling the posterior associated to each dataset using the MCMC technique, we executed four parallel chains with 10,000 steps after the burn-in phase, resulting in 40,000 samples for each posterior distribution.

\begin{figure}[t]
  \centering
  \includegraphics[width=\columnwidth]{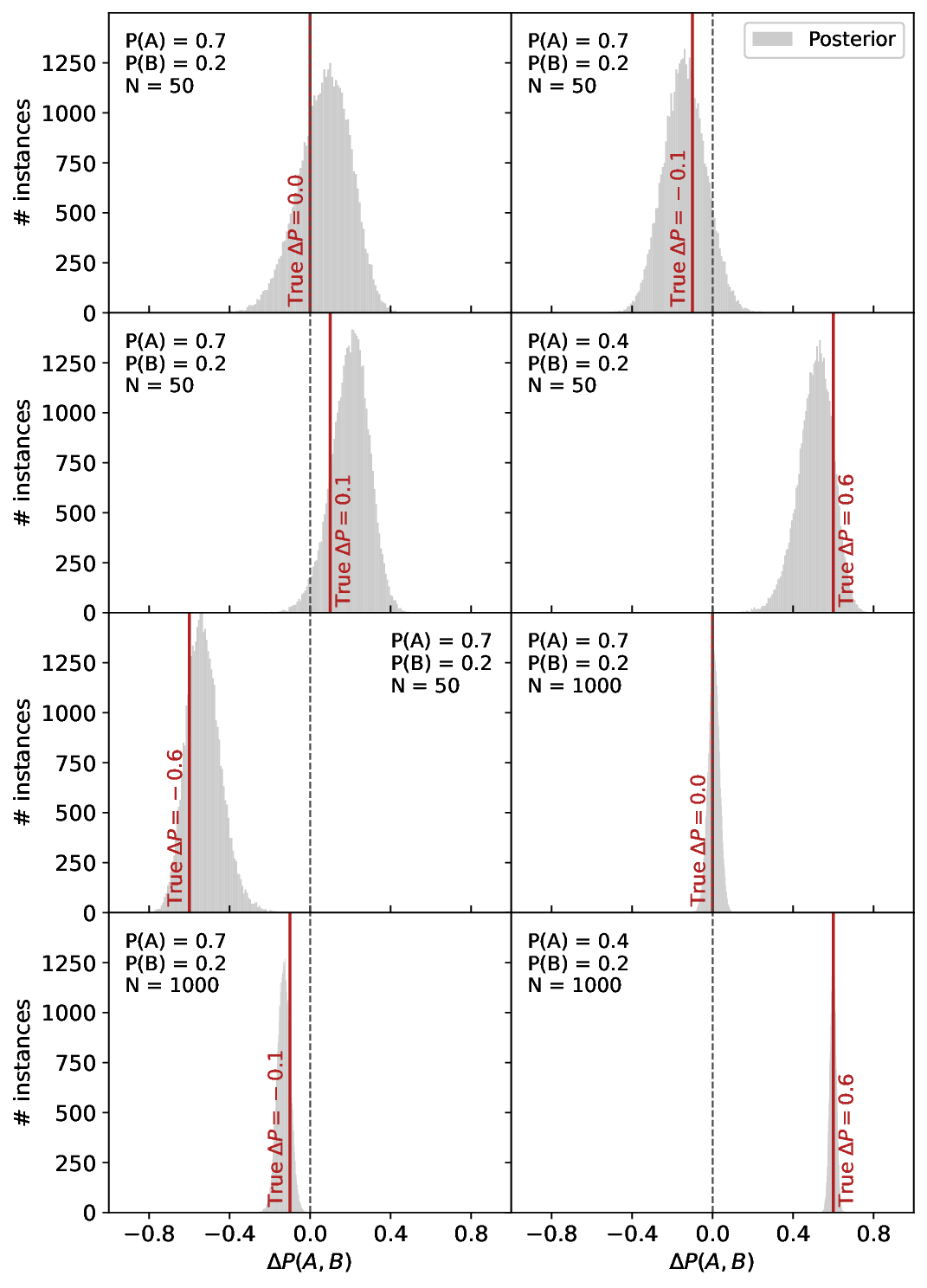}
  \caption{Histograms depicting the sampled $\Delta P(A, B)$ posterior distributions, showcasing results from eight simulations characterized by varying parameters of $\Delta P(A, B)$, $P(A=1)$, $P(B=1)$, and sample size $N$.}
  \Description{The posteriors cover the region around the true probability boost values set for the simulations. When the sample size is larger, the posterior is narrower.}
\label{fig:posterior-dists}
\end{figure}

Figure \ref{fig:posterior-dists} displays the $\Delta P(A, B)$ posteriors for each set of underlying parameters and compares them with the true $\Delta P(A, B)$ values and zero. Each plot's solid red and dashed gray vertical lines represent these values. It is evident that the method is effective in all scenarios, and larger sample sizes result in narrower posteriors, as anticipated.

Figure \ref{fig:posterior-dists} further illustrates how the Bayesian framework simplifies the computation and interpretation of statistical significance and uncertainty of the measure of association (in our case, $\Delta P$). Despite the top-right and bottom-left panels having the same $\langle\Delta P\rangle$, the top-right one exhibits a significantly larger uncertainty, compromising its statistical significance (its $p$-value is approximately 10\%). 

\section{Application to sentiment analysis}
\label{sec:application}

Emotions serve as intricate responses to environmental stimuli, composed of multiple elements such as physiological shifts, cognitive adjustments, and behavioral manifestations \cite{Schwarz2002}.

Moreover, an emotional response is also accompanied by a subjective experience. This experience encompasses the conscious interpretation of the event, intertwined with beliefs, expectations and the specific context in which the emotion occurs, which are referred to as cognitive labels. Generally, emotions operate as adaptive mechanisms, priming the organism for particular courses of action. Importantly, they also have a pronounced sociocultural dimension that affects how they are expressed, interpreted, and utilized within distinct social environments.

Many emotions have important roles in social situations. For example, feelings like joy or thankfulness can help build strong relationships, while emotions like anger or sadness can send strong messages that help groups take action or offer support. Some emotions can even make social conflicts more intense. This is especially true in today's digital world. Social media platforms like X (formerly Twitter) and Facebook have become key places to study how people show and share emotions. Studies have found that the emotions people express online can change the kinds of messages that get shared and even the way people interact with each other \cite{Chmiel2011}.

Understanding emotions on social media is really helpful for several reasons. First, it tells us how people communicate feelings when they're not face-to-face, which affects everything from friendships to political discussions. Second, it helps us understand why some messages go viral or why a group's mood might suddenly change. Lastly, looking closely at how people express emotions online can teach us a lot about how feelings are used to achieve social goals, like building online communities or influencing public opinion \cite{Stieglitz2013}.

While emotions may initially appear as discrete entities with unique physiological and expressive attributes, they often exhibit underlying similarities. These intricate connections can be understood through two prominent frameworks in emotional research: dimensional and hierarchical models. Dimensional models, such as the circumplex model, arrange emotions based on two main axes—valence and arousal. According to this framework, emotions with similar valence or arousal levels are more likely to co-occur \cite{Russell1980}. Hierarchical models, conversely, categorize emotions as primary, secondary, or tertiary, depending on their developmental onset and overall complexity. In this schema, primary emotions serve as the foundation for secondary and tertiary emotions, which are closely related and thus often experienced together \cite{Shaver1987}.

While emotions often manifest as distinct psychological and physiological experiences, they are interconnected in ways that are shaped by both cognitive processes and underlying physiological mechanisms. Given the frameworks of dimensional and hierarchical models of emotion, it is plausible to assume that emotions sharing similar valence or hierarchical attributes may co-occur. Alternatively, one emotion may be interpreted as another if they share these attributes. This application aims to analyze the relational occurrences of different emotions within the realm of social media. Specifically, we seek to understand how these emotions are used in social media contexts, and to assess the explanatory power of both dimensional and hierarchical models for our findings.

\subsection{Data collection}

We utilized Tweepy\footnote{\url{https://www.tweepy.org/}}, a Python library for accessing the Twitter API, in a script designed to collect random tweets in Portuguese. Leveraging the \texttt{filter} method of its (now deprecated) \texttt{Stream} class, Tweepy interacted with Twitter's official API, fetching a random sample of real-time tweets that met the specified user requirements. The only requirement we imposed was on the \texttt{languages} parameter, set to \texttt{['pt']}, ensuring the collection of tweets in Portuguese. Retweets were excluded to retain only original content.

To enhance the representativeness and diversity of the sampled content, we distributed our collection process from the 12th to the 20th of March 2022, covering weekdays, weekends, and various times of the day (morning, afternoon, and night). In total, we gathered 5,000 tweets.

\subsection{Text annotation}

We formed one team comprised of three psychology undergraduate students as annotators. The team was assigned the task of classifying tweets based on emotional themes, drawing from a list of 30 categories influenced by a literature of emotions in Portuguese \cite{Cortiz2021}. This list was initially developed through clustering analysis of annotated datasets of diverse natures \cite{Cowen2021}, and it was subsequently expanded to encompass the categories LOVE, AMUSEMENT, SCHADENFREUDE, and NEUTRAL, which are prevalent in Brazilian culture. 

Each annotator worked independently, with the freedom to label a tweet under multiple categories. The annotators were instructed to label all identified emotions in the same tweet. For a tweet to officially fall under a specific category, a consensus of at least two annotators was required. Tweets that only received a single annotator's endorsement for a category were not categorized under that label. Tweets without any consensus were excluded from the dataset. After removing these tweets and those not evaluated by all three annotators, we retained a dataset of 4,613 posts.

\subsection{Results}

Since annotators could assign multiple emotions to each tweet, we treated the identification of each emotion as an independent binary variable without any inter-variable constraints. Since we utilized 30 emotional categories in our study, there are $(30 \cdot 29) / 2 = 435$ pairs of variables for assessing dependence. To address the issue of multiple comparisons and mitigate the risk of false positives, we implemented the Bonferroni correction to the 2\% significance level we would adopt in a single test. Consequently, we only considered pairs of variables with $p$-values below $4.6 \cdot 10^{-5}$ as indicating statistically significant dependencies.

\begin{figure}[t]
  \centering
  \includegraphics[width=\columnwidth]{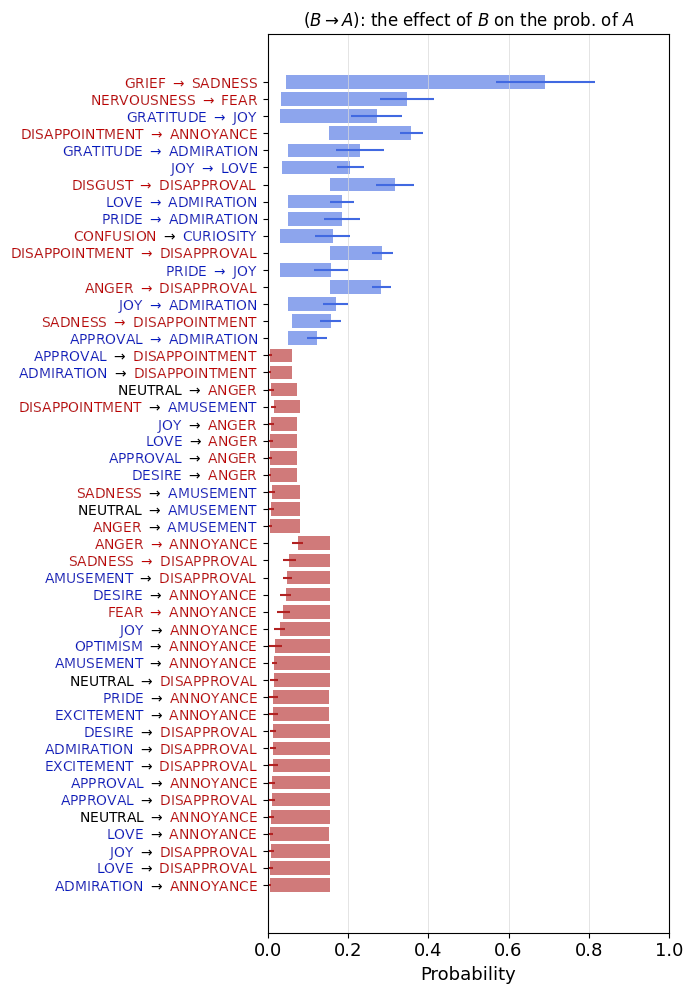}
  \caption{Probability boosts on sentiment $A$ detection given that another sentiment $B$ was already detected (added value). Each bar starts at $P(A=1)$ and ends at $P(A=1|B=1)$. An error bar representing the standard deviation of $P(A=1|B=1)$ features at the end of each bar. The length of each bar represents $\Delta P(A, B)$, and the bar is blue when $\Delta P(A, B) > 0$, and red otherwise. We only included emotion pairs with statistically significant dependence and the $A\leftrightarrow B$ permutation in each emotion pair with the highest $|\Delta P|$.}
  \Description{Horizontal bar plot for each one of the 48 pairs of dependent emotions. The detection of grief boosts the probability for identifying sadness by almost 70 percent. The boosts on other emotions are smaller. Fear gets boosted by 30 percent if nervousness is already identified. Some emotions are almost dichotomic: if one is identified, the chance of detecting the other goes is compatible with zero.}
\label{fig:sentiment-boosts}
\end{figure}

We generated posterior estimates using four parallel MCMC chains, each comprising 10,000 steps after the burn-in phase, resulting in 40,000 samples for every pair of emotions. Among the 435 pairs assessed, we identified 48 as dependent. Figure \ref{fig:sentiment-boosts} illustrates the estimates for $P(A=1)$, $P(A=1|B=1)$, and $\Delta P(A,B)$ for these pairs of emotions. As explained in Section \ref{sec:derivation}, these measures are asymmetrical under the permutation of emotions $A$ and $B$, so we would need $2\cdot 48 = 96$ bars to represent all values in the same plot. To unclutter the plot, we chose to display only the permutation in each pair with the highest $|\Delta P|$.\footnote{To simplify the notation, we will use $\Delta P$ to denote the point estimate $\langle \Delta P \rangle$. It should be clear from context when we are referring to the actual variable.}

The figure reveals several exciting relationships. In general, emotions that co-occur in a tweet tend to share the same polarity, be it positive, negative, or neutral, affirming the validity of such broad sentiment classifications.  To facilitate visualization, the polarity was emphasized by coloring the emotions' names on the vertical axis: blue text for positive, red for negative, and black for neutral. An interesting exception is the notable increase from 3\% to 16\% in the probability of detecting curiosity when confusion is already present. Nonetheless, this association between both emotions remains sensible.

Furthermore, as depicted in Fig. \ref{fig:sentiment-boosts}, there are instances where the presence of one emotion reduces the likelihood of finding the other in the same tweet. In most cases, this reduction drives the probability close to zero, implying that the emotions in these pairs are mutually exclusive. Notably, such exclusivity only occurs between emotions of differing polarities.

The negative boosts that do not drive the probabilities to zero reveal an intriguing finding: while not mutually exclusive, certain negative emotions tend to repel each other. In other words, they appear together less frequently than what chance would suggest. Notable examples include anger and annoyance, as well as disapproval and sadness. This observation might indicate distinct modes of adverse reaction.

Conversely, there are mixed-polarity pairs that, despite displaying repulsion, are not exclusive, meaning they occasionally co-occur. Examples of such pairs are disapproval and amusement, as well as annoyance and desire.

\begin{figure}[ht]
  \centering
  \includegraphics[width=0.7\columnwidth]{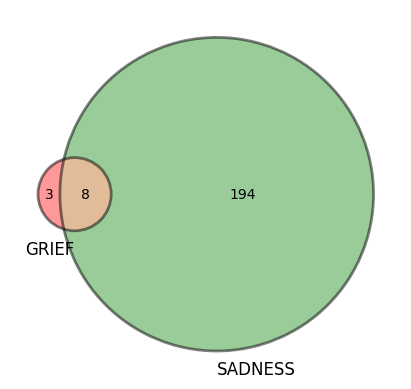}
  \caption{Venn diagram depicting the number of tweets annotated with grief and/or sadness.}
  \Description{194 tweets were annotated with sadness only, 3 with grief only and 8 with both. Since the total number of tweets containing grief is 11, the circle representing these tweets is almost completely inside the sadness circle.}
\label{fig:venn-grief}
\end{figure}

The significant boost that grief applies to sadness is striking: when grief is present, the probability of encountering sadness in a tweet surges from 4\% to 69\%. This substantial increase, which is not as pronounced in the opposite direction, signifies a hierarchical relationship between these two emotions. Grief is closely related to sadness, almost resembling a subclass of it. The Venn diagram in Fig. \ref{fig:venn-grief} visually represents this relationship, illustrating the number of tweets annotated with these two emotions. Out of the 11 tweets annotated with grief, a substantial eight were also annotated with sadness, an emotion detected in a significantly larger set containing 202 instances. The extensive overlap between the sets and their size disparities explain why detecting grief amplifies the likelihood of finding sadness by 65 percentage points (p.p.), whereas identifying sadness elevates the chance of detecting grief by only four p.p.

\begin{figure}[ht]
  \centering
  \includegraphics[width=\columnwidth]{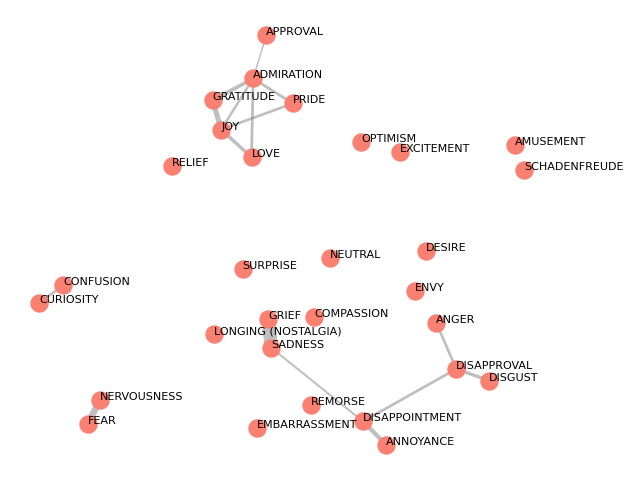}
  \caption{Graph of the relationships between emotions, represented as red nodes. Statistically significant positive probability boosts (added values) are represented by gray connections (graph edges). The width of the connections are proportional to the highest boost in each pair. The nodes' positions were set with the tSNE technique.}
  \Description{A graph with 30 nodes and 16 edges. Positive emotions are clustered on the top, while negative emotions are clustered at the bottom. }
\label{fig:emo-graph}
\end{figure}

Lastly, Fig. \ref{fig:emo-graph} presents an overview of the relationships between all emotions. We clustered the emotions using the added value $\Delta P$ with the highest modulus in each pair as a proximity measure. Specifically, we computed pairwise distances $\delta(A,B)$ between emotions using the formula:

\begin{equation}
\delta(B,A) \equiv 0.01 + \mathrm{max}(\Delta P) - \mathrm{max}[\Delta P (A,B), \Delta P (B,A)], 
\label{eq:distances}
\end{equation}

We employed this measure in the t-distributed stochastic neighbor embedding (t-SNE) \cite{Hinton2002} to project the emotions onto a 2D plane. In the equation above, $\mathrm{max}(\Delta P)$ represents the highest added value among all emotion pairs.

In Fig. \ref{fig:emo-graph}, emotions $A$ and $B$ were connected by a gray line whose width is proportional to $\mathrm{max}[\Delta P (A, B), \Delta P (B, A)]$ when $\Delta P$ was positive and statistically significant. Apart from the clear separation between positive emotions at the top and negative emotions at the bottom, we can identify four clusters of connected emotions and some disconnected ones.

\section{Summary}

This paper presented a Bayesian framework for assessing the degree of association between categorical variables. This framework utilizes the formal definition of variable independence, expressed as $P(A|B) = P(A)$, and was implemented using Markov Chain Monte Carlo (MCMC) techniques. An essential advantage of this method lies in its ability to generate a posterior distribution for our measure of association, providing nuanced insights into the relationship between the two variables under examination and allowing for the computation of confidence intervals and statistical significance for the association measure.

Another distinguishing feature of this method, setting it apart from established statistical tests, is its independence from prior knowledge of the total occurrences for any category. This attribute makes it well-suited for many applications that collect data from the Web, including tasks like sentiment and discourse analysis on a sample of posts from social media. On the downside, the proposed method is much slower than traditional methods that rely on point estimates of association, making it best suited for datasets with a small number of candidate association rules (a few hundred, at most).

There are avenues for further exploration in this research. Firstly, our current prior assumes equal probability levels for detecting and non-detecting any given emotion among a set of 30 possibilities. Enhancing the method might involve adjusting this prior to reflect a higher probability of non-detection.

Additionally, future research could focus on finding an analytical derivation for the posterior of $\Delta P(A, B)$. If existent, the derivation could significantly enhance the efficiency of estimating the degree of association.

In our Twitter-based analysis, we observed consistent co-occurrences of specific emotions in tweets, providing empirical evidence for the predisposition of certain emotions to appear together. These emotional pairings can be interpreted through multiple theoretical lenses. According to the circumplex model \cite{Russell1999}, the similar valence of frequently co-occurring emotions suggests that these combinations are not arbitrary but are likely to be shaped by inherent emotional dimensions. Take, for example, the pairing of ``grief'' and ``sadness'', both of which have negative valence, thereby aligning closely within the circumplex framework.

The hierarchical models of emotion offer another insightful perspective \cite{Shaver1987,Zelenski2000}. In our data, we noted pairings like ``nervousness'' and ``fear,'' where the former, a secondary emotion, commonly appears with the latter, a primary emotion. This finding underscores the hierarchical organization of these emotional states, supporting the notion that primary emotions serve as building blocks for secondary and tertiary emotions.

In addition to these theoretical implications, our findings suggest a practical dimension related to social communication. Emotions play pivotal roles in social bonding, communication, and even conflict \cite{Keltner1999}. The frequent co-expression of emotions sharing similar valence may enrich Twitter's emotional discourse, thereby contributing to more effective communication, message dissemination, and social interaction. This observation aligns with \cite{Zelenski2000}, who found that positive emotions coalesce more readily than negative ones, a pattern also reflected in our data set.

Conclusively, our findings portray the co-occurring emotions in tweets as a complex interplay influenced by both dimensional and hierarchical frameworks, as well as by the social communicative needs of Twitter users. Future research is encouraged to delve into the repercussions of these emotional pairings on aspects like reader engagement, message virality, and the emotional well-being of the Twitter community.

\section*{Acknowledgements}
\label{sec:acknowledgements}

The following authors acknowledge the financial support from the respective Brazilian funding agencies: Ana Luísa Freitas, National Council for Scientific and Technological Development - CNPq (grant 371611/2023-7); Mateus Silvestrin, São Paulo Research Foundation - FAPESP (grant 2021/14866-6); Letícia Morello, São Paulo Research Foundation - FAPESP; Paulo Boggio, Coordination for the Improvement of Higher Education Personnel - CAPES (grant 88887.310255/2018–00), National Council for Scientific and Technological Development grant - CNPq (309905/2019-2), and CNPq - INCT (National Institute of Science and Technology on Social and Affective Neuroscience, grant n. 406463/2022-0); and Fernanda Pantaleão, São Paulo Research Foundation - FAPESP (grant 2022/05313-6).

\bibliographystyle{ACM-Reference-Format}
\balance
\bibliography{main}

\end{document}